\documentclass[seceq,supplement]{ptptex}

\usepackage{graphicx}
\usepackage{wrapft}

\title{
Toward understanding the lattice QCD results\\
 from the strong coupling analysis%
}

\author{
Kenji \textsc{Fukushima}%
}

\inst{
Center for Theoretical Physics, Massachusetts Institute of Technology,\\
Cambridge, Massachusetts 02139, USA\\
and\\
Department of Physics, University of Tokyo,\\
7-3-1 Hongo, Bunkyo-ku, Tokyo 113-0033, Japan%
}

\abst{
This is a review of strong coupling approaches to grasp the nature of
the phase transition in finite temperature and density QCD. We
commence with classics of the center symmetry and the Polyakov loop in
pure gauge theories. The effective action derived in the strong
coupling limit gives qualitatively plausible behavior for the
deconfinement transition at finite temperature. We can apply the
strong coupling analysis to describe the chiral phase transition with
the help of large dimensional expansion. In order to make this article
as self-contained as possible, we elucidate the computational
procedure and its physical meaning in detail. Also the relation
between the Polyakov loop behavior and the chiral dynamics, the phase
structure of two-color QCD, and future possibilities to be pursued in
the strong coupling analysis are discussed.
}

\begin{document}
\maketitle


\section{Objectives}
The aim we are trying to achieve here is to understand the phase
structure of Quantum Chromodynamics (QCD) in a medium at finite
temperature and baryon density. It is known that QCD would undergo
phase transitions between different states of matter, namely, the
hadronic phase, the quark-gluon plasma (color-deconfined or chiral
symmetry restored phase) \cite{ris03}, the color-super conductivity or
diquark super-fluidity \cite{raj00}, and some other exotic states like
pion and kaon condensations, ferromagnetic spin alignment, etc.

The lattice QCD simulation is one of the most prosperous implements to
draw the \textit{right} answer among various possibilities
\cite{lae03}. The phase structure of QCD or QCD-like theories have
been intensely investigated by the lattice simulation at finite
temperature. In many cases the physical quantities of interest in the
lattice calculation are the \textit{Polyakov loop} and the
\textit{chiral condensate}. The Polyakov loop is supposed to be an
order parameter characterizing the deconfinement transition
\cite{pol78,sve86}, as explained later. The chiral condensate is a
familiar quantity serving as an order parameter for the chiral phase
transition.

Even though we could extract the right answer from the lattice QCD
results, we cannot reach a deep understanding of the QCD phase
structure until we correctly comprehend, interpret and explain the
numerical outputs. In a sense the lattice results may be regarded as
experimental facts from which we should draw the physical meaning. For
this purpose the effective model approach would be useful to abstract
the essence. Also, on a technical level, model studies are
indispensable particularly at finite baryon density because the
lattice simulation cannot avoid suffering from the notorious negative
sign problem of the Dirac determinant.

There are many successful effective models so far, such as the
Nambu--Jona-Lasinio model \cite{nam61}, the linear sigma model
\cite{lee72}, the chiral random matrix model \cite{shu93}, the
instanton liquid model \cite{sch98}, the Polyakov loop model
\cite{pis00}, and so on. In this review, we shall draw attention to
another effective model description based on the strong coupling
expansion. Even though the effective action given at strong coupling
has no guarantee to lead to quantitative agreement with physical
observables in the weak coupling regime, the strong coupling approach
has advantages over other models as follows: 1) Because the effective
model derived in the strong coupling expansion has an obvious
connection to the fundamental theory, the result could be exact, in
principle, in the limit of strong coupling. 2) The strong coupling
expansion is the most natural approach on the lattice. Since it is
formulated on the lattice, the correspondence to the lattice QCD
simulation is transparent. 3) The strong coupling analysis is almost
a unique method in which the chiral and the Polyakov loop dynamics
are taken into account on equal footing. 4) The model parameters are
few, namely, the lattice spacing $a$ and the current quark mass $m_q$.
To our surprise, we can reach qualitative and even quantitative
agreement with physical observables with these two parameters.

\begin{wrapfigure}[17]{l}{65mm}
\begin{center}
\includegraphics[width=55mm]{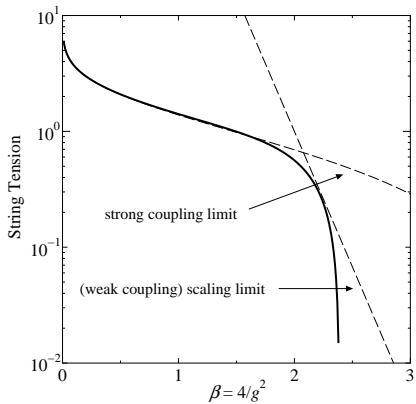}
\caption{String tension in unit of the lattice spacing calculated up
to twelfth order \cite{mun81}. Two dashed curves represent the extreme
cases of strong and weak coupling limits.}
\label{fig:tension}
\end{center}
\end{wrapfigure}

It should be noted here that we must be cautious about the continuity
between the strong and weak coupling regions. If there is a phase
transition at a certain value of the coupling constant, the strong
coupling physics would not give any information on the weak coupling
physics in reality. Although it is not resolved whether the strong
coupling world is smoothly connected to the real world without any
singularity, the string tension of the SU(2) gauge theory calculated
up to twelfth order of the strong coupling expansion \cite{mun81}
suggests that physics continuously changes from the strong coupling
limit toward the scaling limit in the perturbative regime (see
Fig.~\ref{fig:tension}).


\section{Deconfinement and Chiral Restoration}
The deconfinement phase transition is well-defined only in the pure
gauge theory (gauge theory without quarks in the fundamental
representation in color space). In this case we can determine whether
the vacuum is a confined one or not by putting a test static
color-charge. If the free energy gain $F_{\text{q}}$ diverges, any
color-charge isolation is not allowed in this system so that
confinement is realized. [If a theory contains dynamical quarks,
$F_{\text{q}}$ is always finite due to string breaking effects. It is
difficult then to characterize confinement by thermodynamic properties
\cite{fuk03a}.]

$F_{\text{q}}$ can be calculated from the partition function with a
test charge carried by a static quark $q^\alpha(\vec{r})$, which can
be explicitly written down as \cite{sve82,sve86}
\begin{equation}
 F_{\text{q}}(\vec{r})= -T\ln\int\mathcal{D}A_i
  \sum_{\alpha=1}^{N_{\text{c}}}\langle A_i;q^\alpha(\vec{r})|\,
  \mathrm{e}^{-\beta H}\mathcal{P}|A_i;q^\alpha(\vec{r})\rangle
\end{equation}
in the Weyl gauge ($A_0=0$). $\mathcal{P}$ is the projection operator
onto the physical state which should satisfy the Gauss law, that is;
\begin{equation}
 \biggl(D_i\Pi_A^{i\alpha}-\mathrm{i}g\Pi_q\frac{\lambda^\alpha}{2}q
  \biggr)|A_i;q^\beta(\vec{r})\rangle = \biggl\{D_i\Pi_A^{i\alpha}
  +g\frac{\lambda^\alpha}{2}\delta^{\alpha\beta}\delta(\vec{x}
  -\vec{r})\biggr\}|A_i\rangle =0
\end{equation}
with the canonical momenta $\Pi_A^{i\alpha}$ and $\Pi_q$ for the gauge
and quark fields respectively. $D_i$ stands for the covariant
derivative. The Gauss law constraint needs a Lagrange multiplier which
can be eventually regarded as $A_4$ (temporal component of the gauge
field in Euclidean space-time) \cite{gro81}. Then in the
imaginary-time (Matsubara) formalism the free energy can be expressed
as
\begin{equation}
 F_{\text{q}}(\vec{r}) \!=\! -T\ln\!\int\!\mathcal{D}A_\mu\,
  \mathrm{e}^{-S[A]}\,\mathrm{Tr_c}\mathcal{T}\exp\biggl[\mathrm{i}g
  \int_0^\beta\mathrm{d}x_4\,A_4(\vec{r},x_4)\biggr]
  \!\equiv\! -T\ln\langle\mathrm{Tr_c}L(\vec{r})\rangle,
\label{eq:def_polyakov}
\end{equation}
where $\mathcal{T}$ is time ordering and $\mathrm{Tr_c}$ is the trace
in color space. The Polyakov loop is denoted by $L$. If we introduce a
pair of test charge and anti-charge in the same way, we can attain the
inter-quark potential, $F_{\mathrm{q\bar{q}}}$, expressed by the
Polyakov loop correlation function. In Table~\ref{tab:pol_sum} the
basics on the Polyakov loop behavior are summarized.

\begin{table}
\begin{center}
{\small
 \begin{tabular}{lll}
\hline
 & confined (disordered) phase & deconfined (ordered) phase\\
\hline\vspace{-2.3mm} \\

free energy & $F_{\text{q}}=\infty$ & $F_{\text{q}}<\infty$ \\
 & $F_{\mathrm{q\bar{q}}}(\vec{r})\sim\sigma|\vec{r}|$ &
   $F_{\mathrm{q\bar{q}}}(\vec{r})
    \sim-\mathrm{e}^{-m|\vec{r}|}/|\vec{r}|$
\vspace{2mm} \\
Polyakov loop & $\langle\mathrm{Tr_c}L\rangle=0$ &
  $\langle\mathrm{Tr_c}L\rangle\neq0$ \\
 & $\langle\mathrm{Tr_c}L^\dagger(0)\mathrm{Tr_c}L(\vec{r})
  \rangle\rightarrow 0$
 & $\langle\mathrm{Tr_c}L^\dagger(0)\mathrm{Tr_c}L(\vec{r})
  \rangle\nrightarrow 0 \quad (|\vec{r}|\rightarrow\infty)$
\vspace{2mm} \\
\hline
 \end{tabular}
}
\end{center}
\caption{Confined and deconfined phases in the pure gauge theory.}
\label{tab:pol_sum}
\end{table}

As is clear in Table~\ref{tab:pol_sum}, the expectation value of the
Polyakov loop serves as an order parameter to identify the
deconfinement transition. In many cases a phase transition is linked
with the spontaneous symmetry breaking. As a matter of fact, the
Polyakov loop behavior is prescribed by the \textit{center symmetry}
\cite{sve82,sve86,gro81}. The center symmetry in the
SU($N_{\text{c}}$) gauge theory is defined by the gauge transformation
satisfying a boundary condition twisted by an element, $z$, of the
center Z($N_{\text{c}}$) \cite{tho78}. The point is that such a gauge
transformation does not affect the periodic boundary condition for the
gauge field, while the Polyakov loop picks up the element $z$. As a
result the expectation value of the Polyakov loop vanishes if the
vacuum preserves the center symmetry. Hence, concerning the
deconfinement transition, the SU($N_{\text{c}}$) gauge theory can be
regarded as a spin system with the global Z($N_{\text{c}}$) symmetry
described in terms of the Polyakov loop ``spin'' variables. The
universality argument tells us the critical properties of the
deconfinement transition which have been confirmed in the lattice
simulations \cite{lae03,sve82,eng81,kog83}.

One might have thought that it should be curious that the disordered
(ordered) phase lies in the lower (higher) temperature side
\cite{smi96}. This seeming contradiction comes from the duality
transformation connecting the gauge theory to a spin system. Going
back to the original paper written by Polyakov \cite{pol78}, we can
find the explicit transformation for the partition function (notation
is slightly changed);
\begin{equation}
 \int\mathcal{D}\phi\sum_n\exp\biggl[-\frac{g^2}{2T}\sum
  n_{x\mu}^2+\sum\mathrm{i}(\nabla n)\phi\biggr] \sim
  \int\mathcal{D}\phi\sum_m\exp\biggl[-\frac{T}{2g^2}\sum(\mathrm{i}
  \nabla \phi-2\pi m)^2\biggr].
\end{equation}
The left-hand-side is the partition function given by Hamiltonian in
the strong coupling limit on the lattice. The electric flux string,
$n_{x\mu}$, dominates over the magnetic fluctuation at strong
coupling. The Gauss law is implemented by the Lagrange multiplier
$\phi$ (that is $A_4$ in Eq.~(\ref{eq:def_polyakov})). The
right-hand-side is derived by Poisson's resummation formula (duality
transformation) up to an overall factor and is nothing but an
approximated form of the \textit{XY} spin model. From the above
expression it can be clearly understood why the effective ``spin''
theory of the Polyakov loop has the inverse temperature.

In the rest of this section we shall briefly summarize the definitions
inherent in the lattice description. The Wilson action of the gauge
field is given by \cite{wil74}
\begin{equation}
 S_{\text{G}}[U]=\frac{2N_{\text{c}}}{g^2}\sum_{x,(\mu,\nu)}
  \biggl\{1-\frac{1}{N_{\text{c}}}\text{Re}\,\mathrm{Tr_c}\,
  U_{\mu\nu}(x)\biggr\}
\label{eq:wilson_act}
\end{equation}
with the plaquettes;
$U_{\mu\nu}(x)=U_\nu^\dagger(x)U_\mu^\dagger(x+\hat{\nu})
U_\nu(x+\hat{\mu})U_\mu(x)$. The coupling constant, $g^2$, only
appears as an overall coefficient, from which the strong coupling
expansion is derived. The Polyakov loop can be equivalently written on
the lattice as
\begin{equation}
 L(\vec{x})=\prod_{x_d=a}^{N_\tau a}U_d(\vec{x},x_d)
\end{equation}
in $d-$dimensional space-time. $N_\tau$ is the temporal extent and
related to the temperature by $T=1/N_\tau a$ with the lattice spacing
$a$.

As for the chiral symmetry, the lattice fermion suffers from the
notorious species doubling problem. Among ideas to avoid having
$2^d$-fold doublers, we shall limit our discussion in this review to
the staggered formalism (Kogut-Susskind fermion \cite{kog75}) that has
advantage in providing a simple description of chiral symmetry. The
staggered fermion $\chi(x)$ is derived from the Dirac fermion
$\psi(x)$ by the following transformation;
\begin{equation}
 \psi(x)=\Gamma(x)\chi(x),\quad
 \bar{\psi}(x)=\bar{\chi}(x)\Gamma^\dagger(x),\quad
 \Gamma(x)=\gamma_1^{x_1}\gamma_2^{x_2}\cdots\gamma_d^{x_d}.
\end{equation}
Then Dirac's gamma matrices are reduced to be unity and we can forget
about the Dirac indices. As a result $2^d$-fold doublers are
diminished by $2^{[d/2]}$. The remains are usually interpreted as
flavor degrees of freedom. Since the flavor number might be sometimes
confusing as it is, we will write $n_{\text{f}}$ to mean the original
flavor number associated with staggered fermions and $N_{\text{f}}$
to mean the flavor number in the continuum limit (i.e.,
$N_{\text{f}}=4n_{\text{f}}$ for $d=4$).

It should be noted that the chiral symmetry is given by a
U($n_{\text{f}}$)$\times$U($n_{\text{f}}$) rotation on alternate
lattice sites. Although the symmetry pattern looks different, the
chiral condensate, $\langle\bar{\chi}\chi\rangle$, serves as an order
parameter for the chiral symmetry breaking;
$\mathrm{U}(n_{\text{f}})\times\mathrm{U}(n_{\text{f}})\to
\mathrm{U}_{\text{V}}(n_{\text{f}})$. The action for the staggered
fermion is given by
\begin{equation}
 S_{\text{F}}[U,\bar{\chi},\chi] = m_q\sum_x \bar{\chi}(x)\chi(x)
  +\frac{1}{2}\!\sum_{x,\mu>0}\eta_\mu(x)\bar{\chi}(x)\bigl\{
  U_\mu(x)\chi(x+\hat{\mu})-U_\mu^\dagger(x-\hat{\mu})
  \chi(x-\hat{\mu})\bigl\}
\label{eq:staggered}
\end{equation}
with $\eta_\mu(x)=(-1)^{x_1+x_2+\cdots x_{\mu-1}}$. The price for
reducing spin indices is that the flavor contents are scattered and
mixed on the adjacent lattice sites.


\section{Deconfinement Transition}
The lattice gauge theory was originally formulated by Wilson
\cite{wil74} with the aim of giving a plain explanation of
color-confinement in the strong coupling limit. The strong coupling
expansion on the lattice is achieved by statistical analog at high
temperature. The formulation has been developed systematically. In
this article we will not go into mathematical and technical details in
general. If necessary, one can consult an exhaustive review written by
Drouffe and Zuber \cite{dro83}.

\begin{wrapfigure}{r}{52mm}
\begin{center}
\includegraphics[width=38mm]{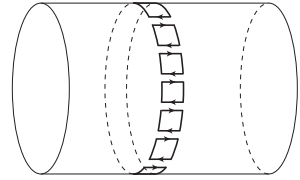}
\caption{Leading-order contribution to the effective action with
respect to the Polyakov loop. The cylinder represents temporal $S^1$
and spatial $R^3$ manifold.}
\label{fig:pol_act}
\end{center}
\end{wrapfigure}

In this section we shall focus on the strong coupling study on the
deconfinement transition for which the order parameter is given by the
Polyakov loop. We can attain the effective action in terms of the
Polyakov loop by integrating over other degrees of freedom. As seen
from (\ref{eq:wilson_act}), the large $g^2$ expansion is
systematically available by the Taylor expansion of exponential of
$S_{\text{G}}$. In the leading order of the strong coupling
expansion the partition function with the Polyakov loop (or $U_d$)
left unintegrated is calculated as (see Fig.~\ref{fig:pol_act})
\begin{align}
 Z_{\text{eff}}^{\text{(G)}}[U_d] &=\int\mathcal{D}U_i\,
  \mathrm{e}^{-S_{\text{G}}[U_i,U_d]}
 =\sum_{m=0}^\infty\frac{1}{m!}\int\mathcal{D}U_i\biggl\{
  \frac{1}{g^2}\sum_{x,(\mu,\nu)}\mathrm{Tr_c}\Bigl(
  U_{\mu\nu}(x)+U_{\mu\nu}^\dagger(x)\Bigr)\biggr\}^m \notag\\
 &=1+\sum_{\vec{x},\hat{j}<d}\mathrm{Tr_c}\prod_{x_d=a}^{N_\tau a}
  U_d(\vec{x},x_d)\,\mathrm{Tr_c}\prod_{x'_d=a}^{N_\tau a}U_d^\dagger
  (\vec{x}+\hat{j}a,x_d')\Bigl(\frac{1}{g^2 N_{\text{c}}}
  \Bigr)^{N_\tau}+\text{(h.c.)}+\cdots.
\end{align}
Note that $1/N_{\text{c}}$ emerges as a result of the group
integration. This expression immediately leads to the effective action
as
\begin{equation}
 S_{\text{eff}}^{\text{(G)}}[L]=-\ln Z_{\text{eff}}^{\text{(G)}}
  =-\mathrm{e}^{-\sigma a/T}\sum_{\text{n.n.}}\mathrm{Tr_c}L(\vec{x})
  \,\mathrm{Tr_c}L^\dagger(\vec{y}),
\label{eq:eff_act}
\end{equation}
where $\text{n.n.}$ is the abbreviation to mean the nearest-neighbor
interaction.  We have used $T=1/N_\tau a$ and the familiar
expression of the string tension in the strong coupling limit;
$\sigma a^2=\ln[g^2N_{\text{c}}]$.

The effective action (\ref{eq:eff_act}) describes the
$(d-1)$-dimensional classical spin system with the spin variable
represented by $\mathrm{Tr_c}L$ and the exchange interaction by
$\mathrm{e}^{-\sigma a/T}$ \cite{kog82}. Although (\ref{eq:eff_act})
looks quite simple, non-trivial complexity can originate from the
matrix nature of the Polyakov loop. In the simplest mean-field
analysis the matrix nature is taken into account by the Haar measure
explicitly, that is the Jacobian associated with the variable
transformation from $A_4$ to $L$ \cite{rei96}. Then the effective
potential can be approximated as
\begin{equation}
 V_{\text{eff}}^{\text{(G)}}[L]/T =-2(d-1)
  \mathrm{e}^{-\sigma a/T}|\mathrm{Tr_c}L|^2-\ln
  \mathcal{M}_{\text{Haar}}[L],
\label{eq:crude_approx}
\end{equation}
where
\begin{equation}
 \ln\mathcal{M}_{\text{Haar}}[L]=
 \begin{cases}
  \ln\bigl[1-|\mathrm{Tr_c}L/2|^2\bigr] & \text{for }N_{\text{c}}=2,\\
  \ln\bigl[1\!-\!6|\mathrm{Tr_c}L/3|^2\!+\!8\text{Re}
  (\mathrm{Tr_c}L/3)^3\!-\!3|\mathrm{Tr_c}L/3|^4\bigr] 
  &\text{for }N_{\text{c}}=3.
 \end{cases}
\end{equation}
The effective potential for $N_{\text{c}}=2$ gives a second-order
phase transition \cite{pol82}, while the Z(3) center symmetry allows a
cubic term so that the phase transition is of first-order for
$N_{\text{c}}=3$ \cite{gro83}. Interestingly enough, it is obvious
from (\ref{eq:crude_approx}) that confinement as well as the
Z($N_{\text{c}}$) nature stem only from the Haar measure, as is
consistent with the arguments that the Haar measure is essentially
important for non-perturbative aspects of QCD \cite{goc93}.

Unfortunately straightforward generalization to $N_{\text{c}}>3$ cases
is not available. Instead, as usually performed in the spin
system, the Weiss approximation works well for arbitrary
$N_{\text{c}}$. In this approximation, we take account of the
fluctuation of the individual Polyakov loop surrounded by a constant
mean field \cite{dro83,kog82,cre83,goc85,fuk03}. This results in a
first-order phase transition for $N_{\text{c}}\ge3$. In particular we
can find the critical coupling
$2(d-1)\mathrm{e}^{-\sigma a/T_{\text{d}}}=0.806$ for
$N_{\text{c}}=3$. Once empirical values, the string tension
$\sigma=(425\,\text{MeV})^2$ and the lattice spacing
$a^{-1}=440\,\text{MeV}$ (see next section), are substituted, the
deconfinement temperature is estimated as
$T_{\text{d}}=204\,\text{MeV}$. One can, otherwise, fit the empirical
value, $T_{\text{d}}=270\,\text{MeV}$, by choosing
$a^{-1}=333\,\text{MeV}$ \cite{fuk03}.

In short summary, the strong coupling expansion yields a simple form
of the effective potential in terms of the Polyakov loop. It leads to
even quantitatively acceptable results as well as qualitative
agreement with gross features of the deconfinement transition. Since
it is easy to handle, the effective potential (\ref{eq:crude_approx})
or more sophisticated one \cite{goc85,fuk03} can also be a useful
ingredient to build an effective model and probe underlying physics in
model studies \cite{fuk03_2}.


\section{Chiral Phase Transition}


\subsection{Spontaneous chiral symmetry breaking at $T=\mu=0$}
Now we shall deal with the system with dynamical quarks. The center
symmetry is lost and the deconfinement transition is obscured then.
Instead, the chiral symmetry plays an important role. In this
subsection we will describe the spontaneous chiral symmetry breaking
at $T=\mu=0$ mainly according to the formalism given by Kawamoto and
Smit \cite{kaw81} and Kluberg-Stern, Morel and Petersson \cite{klu83}.

\begin{figure}
\begin{center}
\includegraphics[width=12cm]{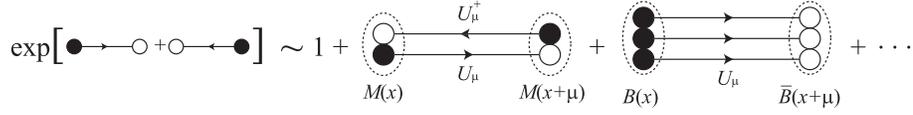}
\caption{Intuitive graphical representation of the $U_\mu$
 integration.}
\label{fig:integ}
\end{center}
\end{figure}

In the $g^2\to\infty$ limit the gauge action can be neglected. We can
accomplish the integration with respect to $U_\mu$ on each lattice
site by expanding the exponential of (\ref{eq:staggered}). Such series
of the Taylor expansion corresponds to the large dimensional
expansion, as explained below. For our present purpose to explain how
it works, we only keep the leading order contributions. Then the
$U_\mu$ integration picks up the mesonic and baryonic (color-singlet)
composites as (see Fig.~\ref{fig:integ})
\begin{align}
 &\int\mathrm{d}U_\mu(x)\;\exp\biggl[-\frac{1}{2}\eta_\mu(x)
  \bar{\chi}(x)U_\mu(x)\chi(x+\hat{\mu})+\frac{1}{2}
  \eta_\mu(x+\hat{\mu})\bar{\chi}(x+\hat{\mu})U_\mu^\dagger(x)
  \chi(x)\biggr] \notag\\
 &=1+\frac{N_{\text{c}}}{2d}\mathrm{Tr_f}M(x)M(x+\hat{\mu})
  +\cdots\notag\\
 &\:-\!(-1)^{N_{\text{c}}(N_{\text{c}}-1)/2}\biggl\{\frac{\eta(x)}
  {2\sqrt{d}}\biggr\}^{N_{\text{c}}}\!\!\mathrm{Tr_f}\bigl\{\bar{B}
  (x\!+\!\hat{\mu})B(x)\!+\!(-1)^{N_{\text{c}}}\bar{B}(x)B(x\!+\!
  \hat{\mu})\bigr\}+\cdots,
\label{eq:trun}
\end{align}
where the mesonic and baryonic composites with flavor indices are
defined by
\begin{equation}
\begin{split}
 & M_{ij}(x) =\frac{1}{N_{\text{c}}}\sqrt{\frac{d}{2}}
  \sum_{\alpha=1}^{N_{\text{c}}}\chi_i^\alpha
  \bar{\chi}_j^\alpha(x), \\
 & B_{i_1 i_2\cdots i_{N_{\text{c}}}}(x) =\frac{d^{N_{\text{c}}/4}}
  {N_{\text{c}}!}
  \sum_{\alpha_1,\dots,\alpha_{N_{\text{c}}}=1}^{N_{\text{c}}}
  \epsilon_{\alpha_1 \alpha_2\cdots \alpha_{N_{\text{c}}}}
  \chi_{i_1}^{\alpha_1}\chi_{i_2}^{\alpha_2}\cdots
  \chi_{i_{N_{\text{c}}}}^{\alpha_{N_{\text{c}}}}(x).
\end{split}
\label{eq:composite}
\end{equation}

Here we can understand where the $1/d$ expansion emerges from. The
large $d$ limit would make sense under the condition that the mesonic
propagation has a finite amplitude. Since the index $\mu$ runs over
$d$ directions, the hopping term $M(x)M(x+\hat{\mu})$ should be
divided by $d$. In other words, the quark fields, $\chi$ and
$\bar{\chi}$, should be normalized by $d^{1/4}$ like
(\ref{eq:composite}). Consequently, the Taylor expansion in terms
of $\chi$ and $\bar{\chi}$ leads to the systematic $1/d$ expansion.

Although the baryonic term may be important when the system has a
non-vanishing baryon chemical potential, we shall drop it from
(\ref{eq:trun}) and examine the spontaneous breaking of chiral
symmetry. [Note that the baryonic term is of higher order in the $1/d$
expansion for $N_{\text{c}}\ge3$.] In order to transform variables
from the quark fields $\chi$ to the meson field $\lambda$, we
linearize the four-fermionic interaction by inserting an auxiliary
field in the following way;
\begin{align}
 &\exp\biggl[N_{\text{c}}\sum_{x,y}
  \mathrm{Tr_f}\biggl\{\frac{1}{2}M(x)V_M(x,y)M(y)+2\bar{m}_q M(x)
  \delta_{xy}\biggr\}\biggr] \notag\\
 =&\int\mathcal{D}\lambda\;\exp\biggl[-N_{\text{c}}\sum_{x,y}
  \mathrm{Tr_f}\biggl\{\frac{1}{2}\lambda(x)V_M(x,y)\lambda(y)-\bigl(
  \lambda(x)V_M(x,y)+2\bar{m}_q \delta_{x,y}\bigr)M(y)\biggr\}\biggr],
\end{align}
where $V_M(x,y)=(1/2d)\sum_\mu(\delta_{x,y+\hat{\mu}}+
 \delta_{x,y-\hat{\mu}})$ represents the mesonic hopping propagation
between the nearest neighbor sites and $\bar{m}_q=m_q/\sqrt{2d}$.

Now that the effective action is linearized in regard to the composite
field $M(x)$, the integration with respect to the quark fields,
$\chi(x)$, is straightforward to be carried out resulting in the
effective action in terms of $\lambda(x)$;
\begin{equation}
 S_{\text{eff}}^{\text{(F)}}[\lambda]=\frac{N_{\text{c}}}{2}\sum_{x,y}
  \mathrm{Tr_f}\lambda(x)V_M(x,y)\lambda(y)-N_{\text{c}}\sum_x
  \mathrm{Tr_f}\ln\Bigl[\sum_y \lambda(y)V(y,x)+2\bar{m}_q\Bigr].
\label{eq:eff_act_f}
\end{equation}
Here $\lambda$ can be regarded as the meson field and its condensate,
which is denoted by $\delta^{ij}\bar{\lambda}$, generates an
additional contribution to the quark mass. The stationary value of
$\bar{\lambda}$ is fixed by the extremal condition of the effective
potential,
\begin{equation}
 V_{\text{eff}}^{\text{(F)}}[\bar{\lambda}] =\frac{N_{\text{c}}
  n_{\text{f}}}{2}\bar{\lambda}^2 -N_{\text{c}} n_{\text{f}}
  \ln\bigl[\bar{\lambda} +2\bar{m}_q\bigr],
\label{eq:eff_chiral}
\end{equation}
to lead to a finite scalar condensate,
$\bar{\lambda}=-\bar{m}_q+\sqrt{1+\bar{m}_q^2}$, which brings about
the spontaneous breaking of chiral symmetry. Actually $\bar{\lambda}$
is directly related to the chiral condensate,
\begin{equation}
 \Psi_q =\langle\bar{\chi}(x)\chi(x)\rangle=
  -\frac{1}{N^d N_{\text{c}}n_{\text{f}}}\cdot
  \frac{\mathrm{d}}{\mathrm{d}m_q}\ln\int\mathcal{D}\lambda\;
  \mathrm{e}^{- S_{\text{F}}[\lambda]},
\end{equation}
by a simple proportionality relation;
$\Psi_q=-\bar{\lambda}\sqrt{2/d}$ (see a reference \cite{fuk03} for
details). We should remark that the logarithmic term in the effective
potential~(\ref{eq:eff_chiral}) makes infinite barrier around the
symmetric vacuum ($\bar{\lambda}=\bar{m}_q=0$) and this logarithmic
singularity alone is responsible for the symmetry breaking in the
present approach.

In the next place let us look into the meson spectrum acquired in this
framework. The kinetic term of the meson fluctuation,
$\delta\lambda(x)$, around the stationary condensate provides us with
the propagator of meson excitations, whose poles correspond to the
dispersion relations physical particles should satisfy. From the
effective action~(\ref{eq:eff_act_f}) we have the kinetic term as
\begin{equation}
 -\frac{N_{\text{c}}}{2}\sum_{x,y}\mathrm{Tr_f}\delta\lambda(x)\Bigl\{
  V_M(x,y)+\frac{1}{(\bar{\lambda}+2\bar{m}_q)^2}V_M^2(x,y)\Bigr\}
  \delta\lambda(y).
\end{equation}
To derive the information on the meson spectrum we transform the meson
propagator into the representation in the momentum space. Then the
inverse propagator is
\begin{align}
 &\frac{N_{\text{c}}}{(\bar{\lambda}+2\bar{m}_q)^2}\sum_{x,y}
  \mathrm{e}^{\mathrm{i}kx}V_M(x,y)\bigl\{(\bar{\lambda}+2\bar{m}_q)^2
  \delta_{y0}+V_M(y,0)\bigr\} \notag\\
 \propto\;\; & \tilde{V}_M(k)\bigl\{(\bar{\lambda}+2\bar{m}_q)^2
  +\tilde{V}_M(k)\bigr\}\;\propto\;\frac{1}{d}\sum_\mu\cos k_\mu
  +(\bar{\lambda}+2\bar{m}_q)^2,
\end{align}
where we used the Fourier transformed hopping propagator in the
momentum space, $\tilde{V}_M(k)$, which is obtained as
\begin{equation}
 \tilde{V}_M(k) =\sum_x\mathrm{e}^{\mathrm{i}kx}V_M(x,0)
  =\frac{1}{d}\sum_\mu\cos k_\mu.
\end{equation}
Now that the meson propagator in the momentum space is known, we can
read the meson masses putting the momentum
$k_\mu=(\vec{0},\mathrm{i}M)+\pi\delta_\mu$ into the propagator. The
last term $\delta_\mu$ takes either $1$ or $0$ representing the
fermion doublers which are absorbed into the flavor degrees of
freedom. From the condition that the inverse of the meson propagator
has zeros, we have the meson spectrum in the present approach as
\begin{equation}
 \cosh M_p = d\{(\bar{\lambda}+2\bar{m}_q)^2-1\}+2p+1.
  \qquad(p=0,1,\dots,d-1)
\end{equation}
$M_0$ and $M_1$ are identified as the masses of the lightest state
belonging to $0^-$ (pion) and mainly $1^-$ ($\rho$ meson),
respectively. $M_2$ and $M_3$ are regarded as $1^+$ ($a_1$ meson) and
$0^+$ ($a_0$ meson) with significant mixtures with $1^-$ and $0^-$
\cite{klu83a}. In order to reproduce the physical values,
$m_\pi=140\;\text{MeV}$ and $m_\rho=780\;\text{MeV}$ \footnote{We
adopt not $m_\rho=770\;\text{MeV}$ but this value faithfully according
to Kluberg-Stern \textit{et al.} \cite{klu83}}, the model parameters
are fixed as
\begin{equation}
 m_q=8\;\text{MeV}, \qquad a^{-1}=440\;\text{MeV},
\end{equation}
which give other physical quantities as listed in Table \ref{tab:klu}.
We would like to conclude this subsection with quoting from
Kluberg-Stern \textit{et al.} -- \textit{``We believe that this is not
a bad result owing to the crudeness of the model.''}

\begin{table}
\begin{center}
\begin{minipage}{7cm}
{\small
\begin{tabular}{ccc}
\hline
 & $g^2=0$ & Physical values\\
\hline \vspace{-2.3mm}\\
 $M_2$ & $910\;\text{MeV}$ & $m_{a_1}=1260\;\text{MeV}$\\
 $M_3$ & $1010\;\text{MeV}$ & $m_{a_0}=980\;\text{MeV}$
\vspace{1mm}\\
 $M_{\text{B}}$ & $1380\;\text{MeV}$ & $m_N=940\;\text{MeV}$
\vspace{1mm}\\
 $\Psi_q$ & $-(390\;\text{MeV})^3$ & $-(240\;\text{MeV})^3$\\
 $f_\pi$ & $190\;\text{MeV}$ & $93\;\text{MeV}$\\
\hline
\end{tabular}
}
\end{minipage}
\begin{minipage}{6cm}
\caption{Hadron spectrum inferred from the strong coupling analysis.
The baryon mass, $M_{\text{B}}$, can be obtained by inserting the
baryonic auxiliary field to (\ref{eq:trun}) and $f_\pi$ from the PCAC
relation.}
\label{tab:klu}
\end{minipage}
\end{center}
\end{table}


\subsection{Chiral restoration at $T\neq0$ and $\mu\neq0$}
Since the $1/d$ expansion works quite well to describe the spontaneous
chiral symmetry breaking at zero temperature, it is interesting to see
what kind of phase structure appears at finite temperature and baryon
density. The extension to the evaluation in a hot and dense medium is
straightforward as found in literatures \cite{goc85,fuk03,dam84}. The
temperature dependence originates from the boundary condition in the
temporal direction and the baryon chemical potential dependence is
introduced by the replacement; $U_d\to\mathrm{e}^\mu U_d$ ($\mu$ being
the \textit{quark} chemical potential) \cite{has83}. These facts
suggest that we should carefully treat the integration along the
temporal direction apart form the others in the spatial
$(d-1)$-dimensional coordinates. We have two strategies to embody it.

One is a physically preferable but technically difficult
method. First, all $U_\mu$ (including $U_d$) are integrated by means
of the $1/d$ expansion. The resultant action is (\ref{eq:trun}) with
the baryonic term in the temporal direction multiplied by
$\mathrm{e}^{N_{\text{c}}\mu}$. Then, after inserting the mesonic and
baryonic auxiliary fields ($\lambda$ and $b$), the integrations
with respect to $\chi$ and $b$ are performed to obtain the effective
potential in terms of $\lambda$. This computational
procedure works well at $T=0$ and $\mu\neq0$. The physical intuition
that the $\mu$ dependence comes from the baryonic excitations would
make sense. Unfortunately for the $T\neq0$ system, however, we have to
take account of not only the baryonic but also the mesonic loops
explicitly and the calculation becomes too complicated.

The other is much simpler to handle. First, $U_d$ are left
unintegrated and the $1/d$ expansion is applied only for the $U_i$
integration. Then, after inserting $\lambda$, the integration with
respect to $\chi$ and $U_d$ are exactly carried out. It should be
mentioned that in this method the baryonic term in (\ref{eq:trun}) is
not necessary to deal with the $\mu$ dependence. In principle, the
$\chi$ integration with unintegrated $U_d$ contains in itself the
thermal excitation of both mesonic and baryonic composite states. The
problem is that the physical meaning is not transparent especially in
the \textit{confined} phase because the thermal excitation is given by
quarks in this method.

Here we will explain the latter method in detail neglecting the
baryonic hopping term. The integration over $U_i(x)$ is done by the
same procedures as in the case at zero temperature. Then the partition
function can be written as
\begin{align}
 &\int\!\mathcal{D}U_d\mathcal{D}\chi\mathcal{D}\bar{\chi}\,
  \exp\biggl[-\frac{1}{2}\sum_x\eta_d(x)\bar{\chi}(x)\bigl\{
  \mathrm{e}^\mu U_d(x)\chi(x\!+\!\hat{d}) - \mathrm{e}^{-\mu}
  U_d^\dagger(x\!-\!\hat{d})\chi(x\!-\!\hat{d})\bigr\} \notag\\
 &\qquad -\frac{ N_{\text{c}}}{2}\sum_{x,y}\lambda(x)V_M(x,y)
  \lambda(y)+ N_{\text{c}}\sum_{x,y}\bigl(\lambda(x)V_M(x,y)
  +2\bar{m}_q \delta_{x,y}\bigr)M(y)\biggr],
\end{align}
where $V_M(x,y)$ represents the hopping propagator only in the spatial
directions. In the mean-field approximation the meson field,
$\lambda(x)$, is simply replaced by constant
$\delta^{ij}\bar{\lambda}$. Then the functional integration with
respect to the quark field, $\chi(x)$, is reduced into the
one-dimensional problem and it can be easily manipulated. After the
$\chi$ integration, the expression is simplified as
\begin{equation}
 \int\mathcal{D}U_d \prod_{\alpha=1}^{N_{\text{c}}}\Bigl[
  2\cosh(N_\tau E)+2\cos(\theta_\alpha-\mathrm{i}\mu)
  \Bigr]^{n_{\text{f}}}
\end{equation}
in the Polyakov loop gauge ($U_d=\text{diag}(\mathrm{e}^{\mathrm{i}
\theta_1},\dots,\mathrm{e}^{\mathrm{i}\theta_{N_{\text{c}}}})$).
For simplicity we will focus on the $n_{\text{f}}=1$ case
($N_{\text{f}}=4$) from now on. Then the integration with respect to
$\chi$ results in the effective potential;
\begin{equation}
 V_{\text{eff}}^{\text{(F)}}[\lambda] =
  \frac{N_{\text{c}}}{2}\bar{\lambda}^2 -\frac{1}{N_\tau}\ln\Biggl[
  2\cosh(N_\tau N_{\text{c}}\mu)+\frac{\sinh[(N_{\text{c}}+1)N_\tau E]}
  {\sinh(N_\tau E)}\Biggr]
\label{eq:res_eff}
\end{equation}
with the definition; $E=\sinh^{-1}(\bar{\lambda}\sqrt{(d-1)/2}+m_q)$.

\begin{figure}
\begin{center}
\begin{minipage}{6cm}
\includegraphics[width=5cm]{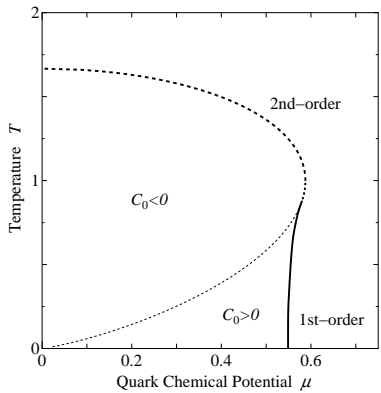}
\end{minipage}
\begin{minipage}{6cm}
\caption{Phase diagram derived from (\ref{eq:res_eff}) with $m_q=0$
and $N_{\text{c}}=3$ in unit of the lattice spacing. $C_0$ is the
potential curvature at $\bar{\lambda}=0$. Thick dashed and solid
curves represent the second and first order boundaries respectively.}
\label{fig:phase}
\end{minipage}
\end{center}
\end{figure}

Now we can draw the phase diagram derived from (\ref{eq:res_eff}) with
$m_q=0$ (chiral limit) and $N_{\text{c}}=3$. As long as $\mu$ is not
large, the chiral phase transition is of second-order. The phase
boundary can be determined by the condition that the potential
curvature around $\bar{\lambda}=0$ (denoted by $C_0$) passes across
zero. When $\mu$ gets larger, the phase transition becomes of
first-order. As a result we reach the phase structure depicted in
Fig.~\ref{fig:phase}. What seems interesting is that in some region of
$\mu$ the chiral phase transition undergoes twice with increasing
temperature unlike the results in other model studies.

\section{Relation between Two Phase Transitions}
In the last section we have seen that the $U_d$ integration is dealt
with separately. Since the Polyakov loop consists of $U_d$, it is easy
to put the Polyakov loop into the formalism. Once we construct an
effective model with the chiral order parameter and the Polyakov loop,
we can investigate the relation between the deconfinement and chiral
phase transitions. An interesting question is why these two
transitions have been observed at the \textit{same} temperature in the
lattice QCD simulations \cite{kog83,fuk86}.

\begin{figure}
\begin{center}
\includegraphics[width=5cm]{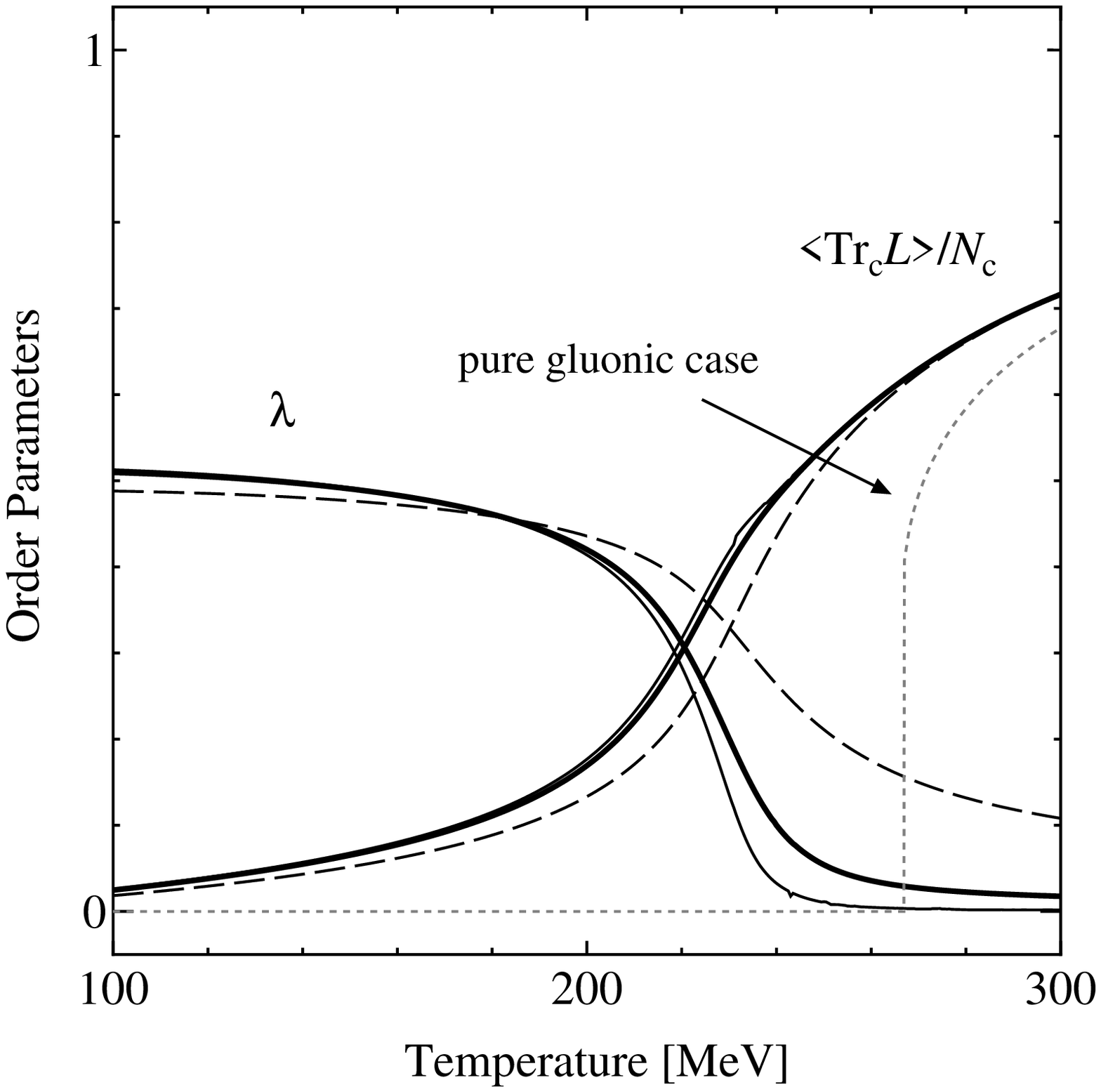}\hspace{1cm}
\includegraphics[width=5cm]{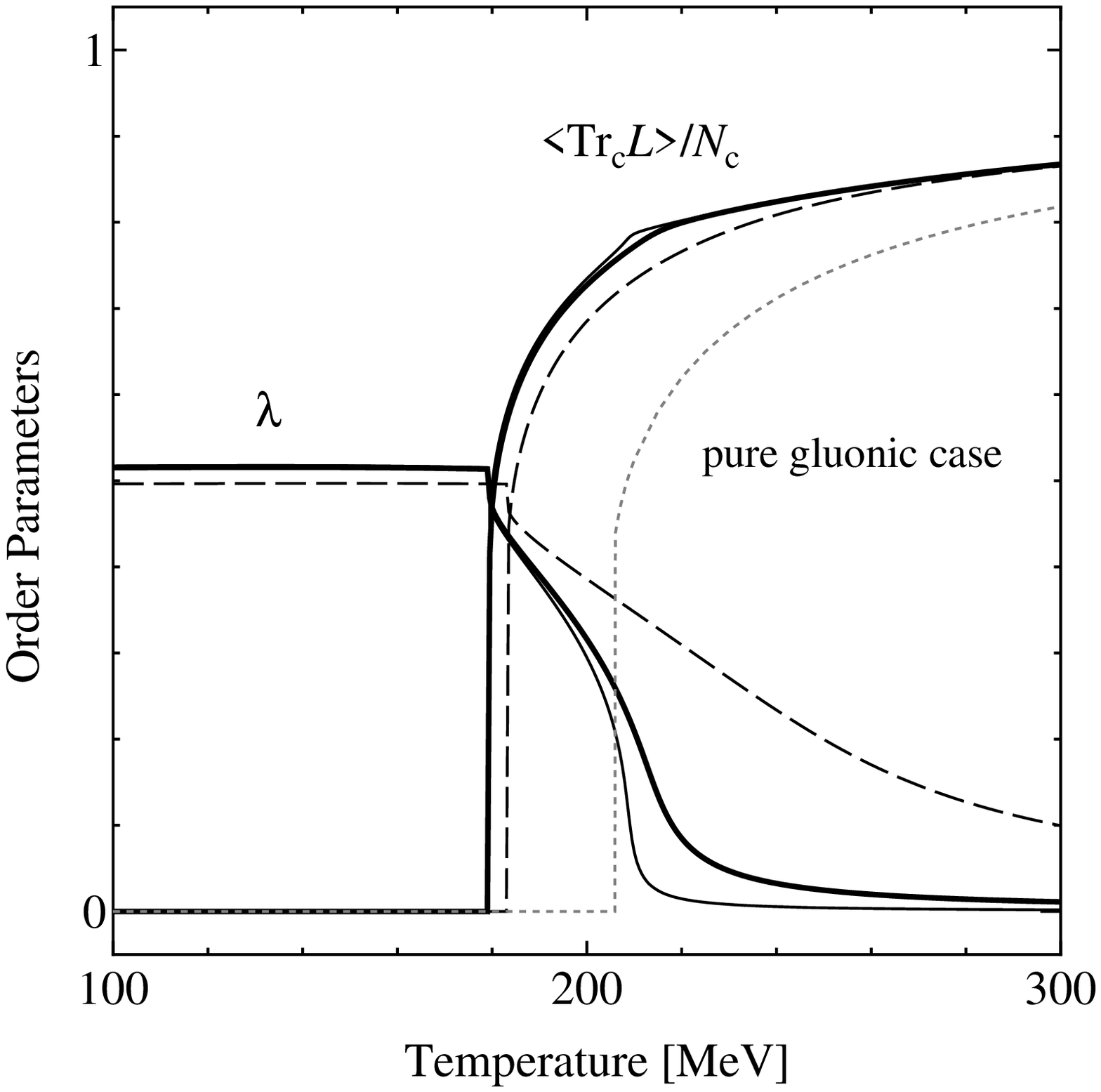}
\end{center}
\caption{The Polyakov loop and the chiral order parameter as functions
of the temperature calculated in the Gocksch-Ogilvie model
\cite{fuk03}. The left figure is for the system with dynamical quarks
in the fundamental representation in color space. The right figure is
for the system with adjoint quarks. Thick curves are for the physical
value of the quark mass and thin solid and dashed curves are for
$m_q=1\;\text{MeV}$ and $m_q=50\;\text{MeV}$ respectively.}
\label{fig:go_res}
\end{figure}

The effect of dynamical quarks on the deconfinement phase transition
was first studied by Green and Karsch \cite{gre84} with the hopping
parameter (heavy quark) expansion. Chiral symmetry was seriously taken
into account by Gocksch and Ogilvie \cite{goc85}, followed by the
finite density extension by Ilgenfritz and Kripfganz \cite{ilg85}. The
Gocksch-Ogilvie model is defined by the effective action; \cite{fuk03}
\begin{equation}
\begin{split}
 & S_{\text{eff}}[L,\lambda] =-\mathrm{e}^{-\sigma a/T}
  \sum_{\text{n.n.}}\mathrm{Tr_c}L(\vec{x})\mathrm{Tr_c}
  L^\dagger(\vec{y}) \\
 &\qquad +\frac{N_{\text{c}}}{2}\sum_{x,y}\lambda(x)V_M(x,y)\lambda(y)
  -\frac{N_{\text{f}}}{4}\sum_{\vec{x}} \mathrm{Tr_c}\ln \biggl[
  \cosh(N_\tau E)+\frac{1}{2}(L+L^\dagger)\biggr].
\end{split}
\label{eq:go_model}
\end{equation}
This can be understood as a combination of (\ref{eq:eff_act}) and
(\ref{eq:res_eff}). The effective potential
$V_{\text{eff}}[l,\bar{\lambda}]$ ($l$ being the expectation value of
the Polyakov loop) obtained from (\ref{eq:go_model}) in
the mean-field approximation has an interesting property, that is,
$V_{\text{eff}}[l=0,\bar{\lambda}]$ inevitably has instability at
$\bar{\lambda}=0$ and thus it leads to a non-vanishing chiral
condensate. Roughly speaking, \textit{the confined vacuum breaks
chiral symmetry at any temperature}. As emphasized in a chiral
effective model study with the Polyakov loop \cite{fuk03_2}, this
property suggests $T_\chi\ge T_d$ (the chiral restoration temperature
is higher than the deconfinement temperature). Remembering that
$T_d\sim270\;\text{MeV}$ without dynamical quarks is higher than
$T_\chi\sim150\;\text{MeV}$, we can expect to have $T_d=T_\chi$ in the
presence of dynamical quarks. As a matter of fact, we can see that in
the Gocksch-Ogilvie model the simultaneous crossovers are attributed
to this mechanism, as shown in the left of Fig.~\ref{fig:go_res}.

\section{QCD-like Theories}


\subsection{QCD with Adjoint Quarks (aQCD)}
Adjoint quarks have the same color structure as gluons and thus never
break the center symmetry. In the lattice simulation of QCD with
adjoint quarks (aQCD), as expected, the deconfinement transition is
found to be of first-order \cite{kar99}. The Gocksch-Ogilvie model has
been generalized to include adjoint quarks and the model study gives a
satisfactory explanation on all the qualitative features
\cite{fuk03}. The right of Fig.~\ref{fig:go_res} is the result in the
Gocksch-Ogilvie model with adjoint quarks.


\subsection{Two-Color QCD}
Two-color QCD (QCD with $N_{\text{c}}=2$) has special properties. As
explained before, the staggered fermion has the global
$\mathrm{U}(n_{\text{f}})\times\mathrm{U}(n_{\text{f}})$ symmetry for
$m_q=0$. For $N_{\mathrm{c}}=2$, the chiral symmetry is graded to
U($2n_{\text{f}}$) \cite{han99,klu83}. This is because the color SU(2)
group is pseudo-real and the action possesses Pauli-G\"{u}rsey's
symmetry \cite{pau57}, which also makes possible the lattice
simulation at finite density.

Introduction of a finite chemical potential $\mu$ or quark mass $m_q$
explicitly breaks the U(2$n_{\text{f}}$) symmetry. Listed in
Table~\ref{tab:su2} are the symmetries realized in various
circumstances and their breaking patterns for $n_{\text{f}}=1$.

\begin{table}
\begin{center}
{\small
\begin{tabular}{|c|c|c|}
\hline
 & $m=0$ & $m\neq0$ \\ \hline
$\mu=0$ & U(2) & $\mathrm{U_V(1)}$ \\
 & broken to U(1) with 3 NG modes & not broken \\ \hline
$\mu\neq0$ & $\mathrm{U_V(1)\times U_A(1)}$ & $\mathrm{U_V(1)}$ \\
 & totally broken with 2 NG modes & totally broken with 1 NG mode \\
\hline
\end{tabular}
\caption{Symmetry realized in the single-component staggered-fermion
action for $N_{\text{c}}=2$. Possible symmetry breaking patterns and
the number of Nambu-Goldstone (NG) modes are also listed.}
\label{tab:su2}
}
\end{center}
\end{table}

\begin{figure}
\begin{center}
\includegraphics[width=5.5cm]{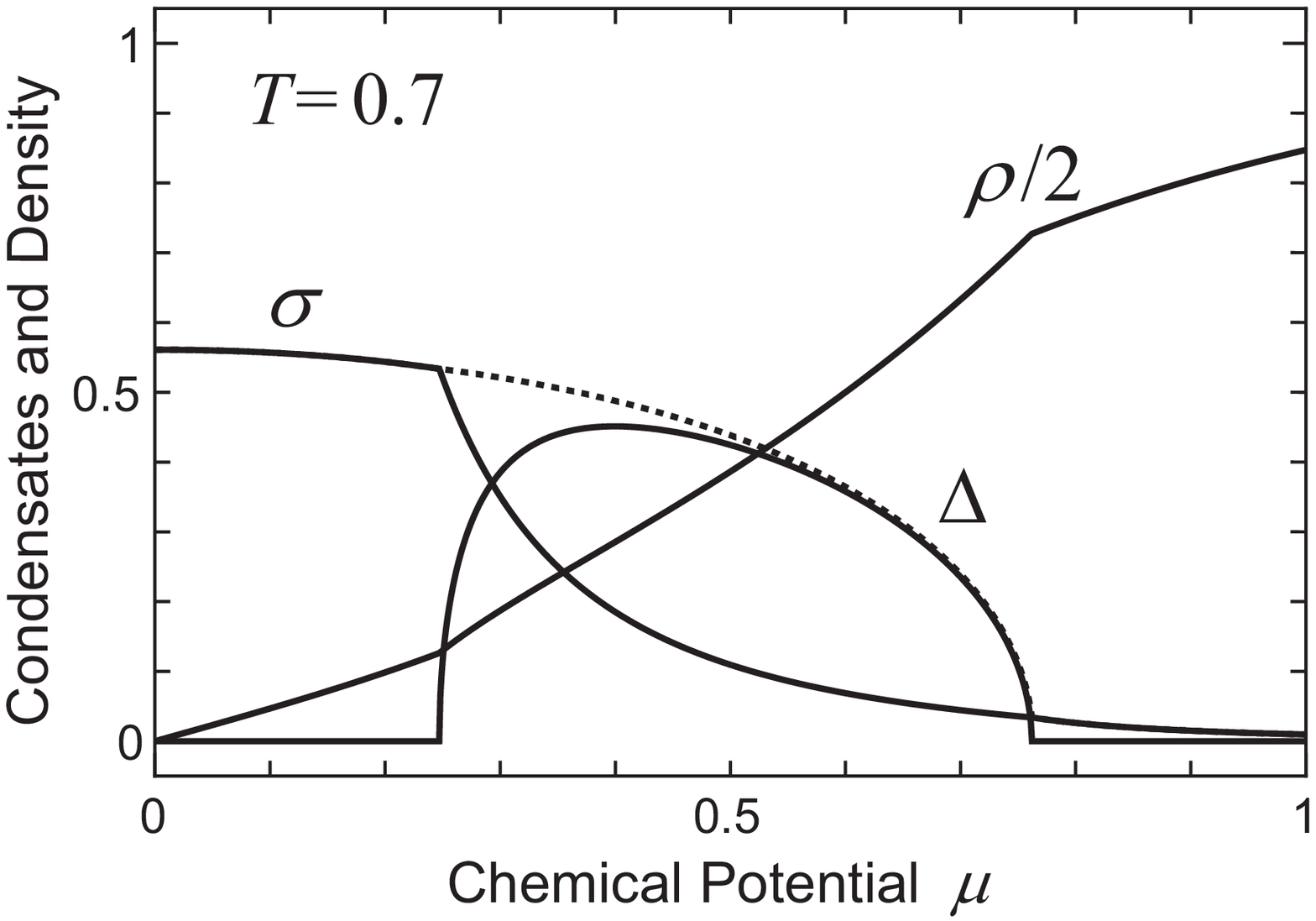}\hspace{7mm}
\includegraphics[width=6.5cm]{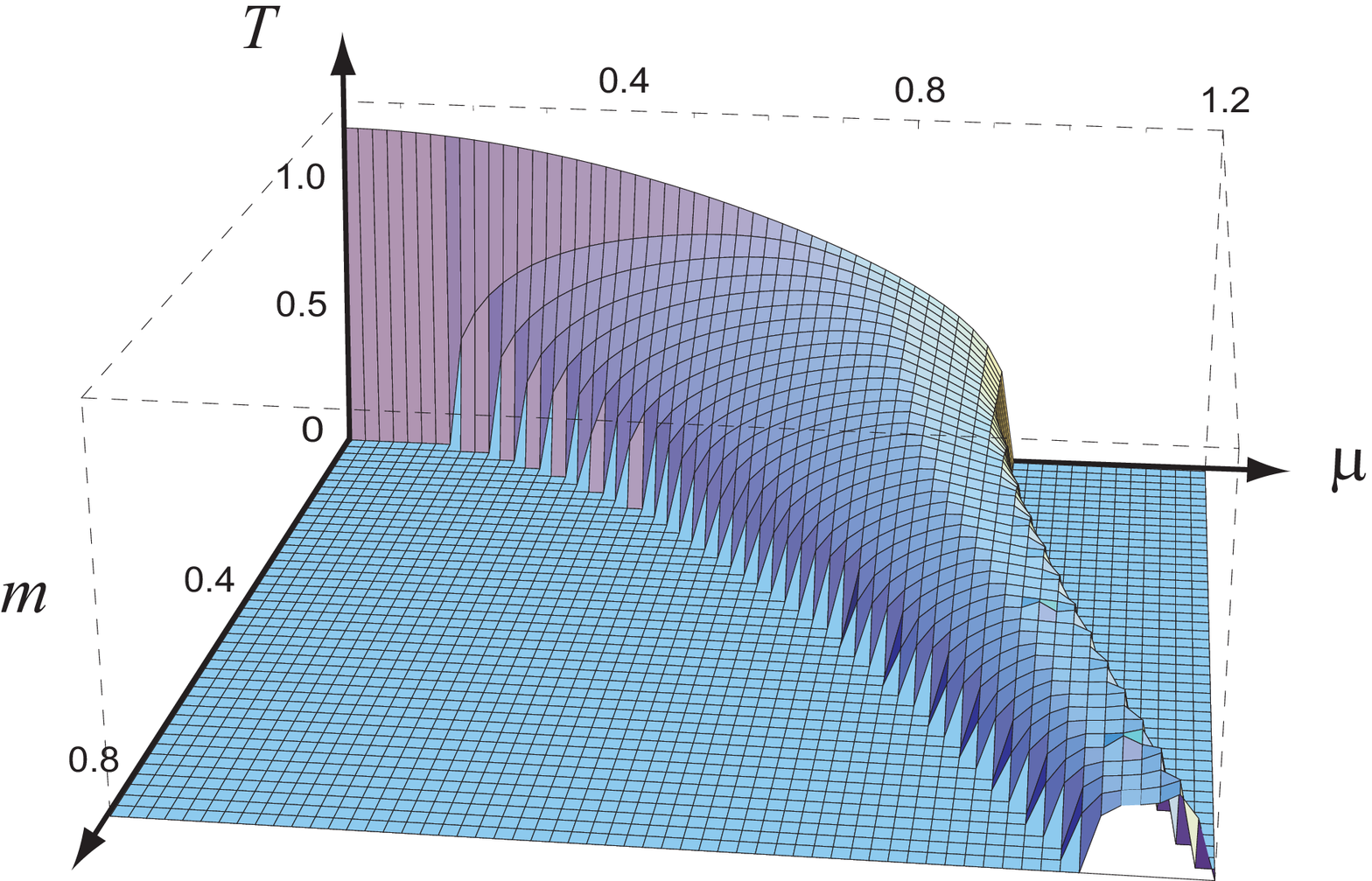}
\end{center}
\caption{Left: Chiral and diquark condensates ($\sigma$ and $\Delta$
respectively) for $m_q=0.02$. Dotted curve represents the magnitude of
condensation; $\sqrt{\sigma^2+\Delta^2}$. The mixing angle between
$\sigma$ and $\Delta$ is controlled by $m_q$. If $m_q=0$ then
$\sigma=0$ and the condensation is saturated by $\Delta$. Right: Phase
boundaries in the 3-dimensional $(m_q,T,\mu)$ space. Below the
critical surface, $\Delta$ takes a non-vanishing value \cite{nis03}.}
\label{fig:su2}
\end{figure}

To our surprise, the chiral condensate,
$\langle\bar{\chi}\chi\rangle$, of two-color QCD vanishes in the
chiral limit ($m_q=0$). Instead, the diquark condensate,
$\langle\chi\chi\rangle$, plays an important role. Note that
$\chi\chi\equiv\epsilon_{\alpha\beta}\chi^\alpha\chi^\beta$ is a
gauge invariant operator for $N_{\text{c}}=2$ and it corresponds to
the baryonic contribution in (\ref{eq:trun}). The phase structure was
first investigated in the strong coupling approach by Dagotto, Karsch
and Moreo \cite{dag86}. The analytical and numerical studies have been
inexhaustively explored within the framework of the strong coupling
limit \cite{dag87,nis03}. The results are consistent with the lattice
simulation \cite{kog01}. The order parameters and the phase diagram
are shown in Fig.~\ref{fig:su2}.

\section{Beyond Understanding}
We have looked over various aspects of the strong coupling approach to
QCD. Since the celebrated paper by Wilson, there are quite a few
contributions to clarify QCD physics in the strong coupling limit.
Nevertheless, there still remain several interesting questions to be
answered in the strong coupling analysis. We shall enumerate some
possibilities here;\vspace{2mm}

1) Glueballs at finite temperature -- In the pure gauge theory, the
Polyakov loop is an order parameter for the deconfinement transition.
However, the physical interpretation and the dynamical description
(real-time evolution) of the Polyakov loop are not clear in Minkowski
space-time. In principle, we can define the deconfinement transition
in terms of physical excitations, namely, glueballs \cite{hat03}. At
zero temperature, glueballs are well described in the strong coupling
expansion \cite{mun81a}. A finite-temperature extension would give us
concrete information on electric and magnetic glueballs near the
deconfinement transition temperature \cite{dat98}.

2) Relation between the deconfinement and chiral phase transitions at
finite density -- Since the lattice QCD simulation has given no
definite answer at finite density, the model study preceding future
lattice simulations at finite density would be especially useful.
Because, unlike the temperature, the baryon density effect explicitly
breaks the center symmetry, it could be expected that the Polyakov
loop behavior should be significantly affected at finite density.

3) Color-super conductivity on the lattice \cite{azc03} -- Although
the lattice simulation is not applicable at finite density, we believe
that the sign problem could be gotten over someday. Then an
interesting and important  problem would be \textit{how to see the
color-super conductivity on the lattice}. Because of Elitzur's theorem
\cite{eli75}, we have to make a gauge invariant order parameter or
choose a certain gauge fixing condition.

4) Phase structure of the system at finite isospin density -- It is
interesting to apply the strong coupling approach to investigate the
phase transition at finite isospin density \cite{son01}, for there
are numerical outputs from the lattice simulation \cite{kog02}. In
order to incorporate the flavor structure the Wilson fermion is more
suitable then. The strong coupling analysis with Wilson fermions
should be developed \cite{ros96}.\vspace{2mm}

Beyond understanding the existing lattice data, we would like to
emphasize that the strong coupling analysis could be a powerful tool
to give some \textit{guideline} or \textit{prediction} to future
progress in the lattice approaches. I hope that this contribution
presented here will provide some clues to open the way for finite
density QCD.

\section*{Acknowledgments}
The author is supported by Research Fellowships of the Japan Society
for the Promotion of Science for Young Scientists. This work is
supported in part by funds provided by the U.S.\ Department of Energy
(D.O.E.) under cooperative research agreement \#DF-FC02-94ER40818.

%

\end{document}